\documentclass[aps,prl,twocolumn,showpacs,superscriptaddress,groupedaddress]{revtex4}  
 \usepackage{verbatim}
 \usepackage{mathtools}
 \begin{document}

	\title{  Average search time bounds in   cue based search strategy}
 	 \author{Vaibhav Wasnik}
 	 \email{wasnik@iitgoa.ac.in}
 	 \affiliation{Indian Institute of Technology, Goa}
	\begin{abstract}
  In this work we   consider the problem of  searches  that utilizes past information gathered during searching, to evaluate the probability distribution of finding the source at each step. We start with a sample strategy where the movement at each step  is in the immediate neighborhood direction, with a probability proportional to the normalized difference in probability of finding the source with the present position source finding probability. We  evaluate a lower bound for the average search time for this strategy.  We  next consider the problem of the lower bound on any strategy that utilities information of the probability distribution evaluated by the searcher at any instant. We derive an expression for the same. Finally we present an analytic  expression for this lower bound in the case of homogeneous diffusion of particles by a source. For a general probability distribution with entropy -$E$, we find that the lowerbound goes as $e^{E/2}$. 
  
	\end{abstract}
	\maketitle
\emph{Introduction:-} Searching for a source that emits particles is a problem that is  quite ubiquitous. We see this all   the way from a bacteria searching for the source of chemoattractants \cite{chemotaxis}, to a robot figuring out the source of a gas leak in a room \cite{robot}. Search time is defined as the time required to find the source by a searcher.  This is similar to the first passage time: the first time the searcher reaches the position occupied by the source. There is a lot of theoretical work done in this area \cite{redner}.  One could classify search strategies into two broad categories. Searches with cues and searches without cues. Searches without cues are reviewed in \cite{benichou}. As has been stated there, searches with cues can also be split up into two kinds. One of them involves chemotatic strategies that assume a sufficient concentration of cues and another category of strategies that involve searching through information coming from sparse cues. Infotaxis \cite{infotaxis} falls in the later category.
 
  A searcher moving through an environment of particles emitted by a source has a history of hits at times $t_1, .., t_n$ at positions $\vec{r}(t_1),..., \vec{r}(t_n)$. These make up the cues that provide all the information from the environment. This information could be utilized in deciding a future direction in many ways. One important quantity that could be measured is the probability of finding the source at any location in space. One could use Bayes theorem to evaluate this as 
  
  \begin{eqnarray}
  P(\vec{r} | \vec{r}(t_1),...,\vec{r}(t_n)) = \frac{P(   \vec{r}(t_1),..., \vec{r}(t_n)) |  \vec{r} ) }{ \sum_x P(   \vec{r}(t_1),..., \vec{r}(t_n)) | \vec{x} ) }
  \end{eqnarray}
here $  P(\vec{r} | \vec{r}(t_1),...,\vec{r}(t_n)$ is the probability of finding the source at position $ \vec{r}$ given hits at positions $  \vec{r}(t_1),...,\vec{r}(t_n) $ and $P(   \vec{r}(t_1),..., \vec{r}(t_n)) | \vec{x} ) $ is probability of hits happening at positions $ \vec{r}(t_1),..., \vec{r}(t_n))$ given the source is at position $ \vec{x}$. Infotaxis \cite{infotaxis} utilizes this probability to evaluate the entropy of the source. The   motion of the searcher at each step is in a direction in which the expected information gain is a maximum.  In \cite{infotaxis} it was  conveyed   that  evaluating the search time analytically for a searcher undergoing Infotaxis, was difficult, given the complexity of the search algorithm. Given this issue, the question arises whether it would be possible to evaluate the search times for a class of cue based searches and any statement be made about certain universal features, such as lower bound on these search times given certain constraints. In this work we  begin with a strategy   that utilizes past cues to evaluate the probability distribution of finding the source at each step and where the searcher at each step moves in the immediate neighborhood, with a probability proportional to the normalized difference in probability of finding the source with the present position source finding probability. We then  attempt to  evaluate a lower bound on the search times in case of homogeous diffusion of particles by a source. We then  consider the problem of the lower bound on any strategy that utilities information of the probability distribution evaluated by the searcher at any instant. We   evaluate an analytical expression for   lower bound in case of homogeneous diffusion of particles by a source. For a general probability distribution with entropy -$E$, we find that the lowerbound goes as $e^{E/2}$. 
	
	\emph{Narrowing   the source:-}	 

 Let us assume that the source emitting particles is located at the origin.   A searcher moving through an environment of particles emitted by a source has a history of hits at times $t_1, .., t_n$ at positions $\vec{r}(t_1),..., \vec{r}(t_n)$. We have,
  \begin{eqnarray}
  P(\vec{r} | \vec{r}(t_1),...,\vec{r}(t_n)) =\nonumber \\
    \frac{ \exp[-\int_0^t P(\vec{r}(t')|\vec{r})] dt'  P(   \vec{r}(t_1),..., \vec{r}(t_n)) |  \vec{r} ) }{ \sum_x \exp[-\int_0^t P(\vec{r}(t')|\vec{x})] dt' P(   \vec{r}(t_1),..., \vec{r}(t_n)) | \vec{x} ) } \nonumber \\
  \end{eqnarray}

Here, $P(   \vec{r}(t_1),..., \vec{r}(t_n)) |  \vec{r} ) $ is the probability of having hits at positions $ \vec{r}(t_1),..., \vec{r}(t_n))$ given the source is at position $r$. The exponentials correspond to no hits happening at the other locations along the trajectory.
Because the hits are independent of each other and can happen at any time.   We could write the above as 
 \begin{eqnarray}
  P(\vec{r} | \vec{r_1},...,\vec{r_n} )\nonumber \\
   = \frac{\exp[-\int_0^t S(\vec{r}(t')|\vec{r}) dt']  S(   \vec{r_1}|\vec{r})   ... S(\vec{r_n} |  \vec{r} )   }{ \sum_x \exp[-\int_0^t S(\vec{r}(t')|\vec{x}) dt' ]S(   \vec{r_1}|\vec{x}) ..S( \vec{r_n}| \vec{x} )  }\nonumber \\
  \end{eqnarray}
  Where above we $\vec{r_1}$..$\vec{r_n}$ are just the positions in space, implying that the probability evaluations are only dependent on the positions in space where hits happen irrespective of the time they happen. The $S( \vec{r_1}|\vec{x} )$ is used above, to imply that the probability of having hits is simply the probability of having particles at location $ \vec{r_1} $  assuming source is at $\vec{x}$.  This assumes that the searcher has an analytical expression for how the particle distribution would be, given the source location. 
  
  The probability that the hits happened at these positions is simply
  \begin{eqnarray}
 e^{-\int_0^t [ S(\vec{r}(t')|0) ]dt'}    S(   \vec{r_1}|\vec{0}) ... S(\vec{r_n} |  \vec{0} )
  \end{eqnarray}
  Hence, the average probability of finding the source at $\vec{r}$ would be 
  
   \begin{eqnarray}
P  (\vec{r})=\sum_{trajectories}\sum_n \frac{1}{n!} \nonumber \\ 
\sum_{ r_1,..r_n}   e^{-\int_0^t [S(\vec{r}(t')|\vec{r}) +S(\vec{r}(t')|0) ]dt'}  S(  \vec{r_1}|\vec{0})  ... S(\vec{r_n}  |  \vec{0} )   &\times&\nonumber \\
 \frac{  S(   \vec{r_1}|\vec{r})   ... S(\vec{r_n} |  \vec{r} )   }{ \sum_x e^{-\int_0^t S(\vec{r}(t')|\vec{x}) dt' } S(   \vec{r_1}|\vec{x}) ...S( \vec{r_n}| \vec{x} )  }\nonumber \\
 \label{bayes_1}
   \end{eqnarray}
   
   It is obvious that if our trajectory took an infinite time we would have the best narrowing of the source location. Hence, the best possible average probability distribution possible telling us the probability of locating the source at position $\vec{r}$ is 
    \begin{eqnarray}
P_\infty(\vec{r})=\sum_{trajectories}\sum_n \frac{1}{n!} \nonumber \\ 
\sum_{ r_1,..r_n}   e^{-\int_0^\infty [S(\vec{r}(t')|\vec{r}) +S(\vec{r}(t')|0) ]dt'}  S(  \vec{r_1}|\vec{0}) ... S(\vec{r_n}  |  \vec{0} ) &\times&\nonumber \\
 \frac{  S(   \vec{r_1}|\vec{r}) ... S(\vec{r_n} |  \vec{r} ) }{ \sum_x e^{-\int_0^\infty S(\vec{r}(t')|\vec{x}) dt' } S(   \vec{r_1}|\vec{x}) ...S( \vec{r_n}| \vec{x} )  }\nonumber \\
 \label{bayes_2}
   \end{eqnarray}

   Let us assume for illustrative purposes that  $ S(   \vec{r_1}|\vec{x}) = S(| \vec{r_1}- \vec{x}| )$. Also, let us assume that $S$ is appreciable only up to a distance $L$ away from the source. One immediately see's from the above expression that because of presence of terms like $  S(  \vec{r_1}|\vec{0})  S(   \vec{r_1}|\vec{r}) $, implies that the average probability distribution of finding the source evaluated above is appreciable over   a distance $2L$, as long as we are considering trajectories of lengths or order larger than $L$ .  This implies that the probability distribution measured by the searcher will not narrow the source better compared to $S$. 
If we consider the limit in which $t\rightarrow 0$ in Eq.\ref{bayes_1}, we can see that the probability distribution measured by the searcher is centered at the searcher position. The measured probability distribution is similarly in general not centered at the position of the source for other values of   $t $.   This implies that the measured probability distribution by the searcher cannot narrow the source better than $S(x)$. 
    
 \emph{Example Strategy:-}   
  Let us consider a search strategy in which the probability to jump to a neighboring location is proportional to the difference in the probability of finding the source from its own location.   
 The  probability for the searcher to jump to the nearest neighbor    $(x+2dx,y)$ on an average  would go  as  $  \beta \; \theta( P(x+2dx,y) - P(x,y) )\frac{ ( P(x+2dx,y) - P(x,y) )}{P(x,y)}  $.   $P(x,y)$ is the average probability of finding the source at position $(x,y)$ that has been evaluated by the searcher using the Baye's theorem as talked in Eq.\ref{bayes_1}.   This could depend on the starting position of the searcher.     $\beta$ is a rate at which this jumps happen and  $\Theta(x)$ is defined as,
	\begin{equation}
  \Theta(x)=\begin{cases}
    1, & \text{if $x>0$}.\\
    0, & \text{otherwise}.
  \end{cases}  
\end{equation}
	  Let us   consider the average time to reach the source from position $(x,y)$ as $T(x,y)$.   As derived in appendix

	 	\begin{eqnarray}
	 0&=&- P(x,y) -    2  \beta \nabla T(x,y) \cdot \nabla P(x,y) -   \beta T(x,y)    \nabla^2 P(x,y) \nonumber \\
	   &+& \alpha \beta  \nabla^2 T(x,y) \nonumber \\
	 \end{eqnarray}

	 For simplicity let probability distribution   have radial symmetry with the source located at $r=0$. Then the above equation simply becomes
	 \begin{eqnarray}
	 0&=& -P(r) -\beta \frac{\partial T(\bar{r})}{\partial r}  \frac{\partial P(\bar{r})}{\partial r}    -  \beta T(\bar{r})      [  \frac{\partial^2 P(\bar{r})}{\partial r^2}  + \frac{1}{r}\frac{\partial  P(\bar{r})}{\partial r } ]   \nonumber\\
	  &+& \alpha \beta P(r)    [  \frac{\partial^2 T(\bar{r})}{\partial r^2}  + \frac{1}{r}\frac{\partial  T(\bar{r})}{\partial r } ]\nonumber \\
	  \end{eqnarray}
 
For $\alpha = 0$ the solution with $T(r=0) = 0$ is 
\begin{eqnarray}
T(r) &=& -\frac{1}{\beta r P'(r)} \int_0^r x P(x)  dx
\end{eqnarray} 
 
 As talked above, the probability distribution $P(x)$ would never be as localized near the source as $S(x)$.  In case we are considering homogeneous diffusion by a source at the origin, in two dimensions, the equilibrium particle concentration at $r$ goes as  $K_0(r/l)$. Hence the lower bound on search time simply is 
  
   \begin{eqnarray}
T(r)> LB(r) &=& -\frac{1}{\beta r K_0'(r/l)} \int_0^r x K(x/l)  dx
\end{eqnarray}

\begin{figure}
  \includegraphics[scale=.75]{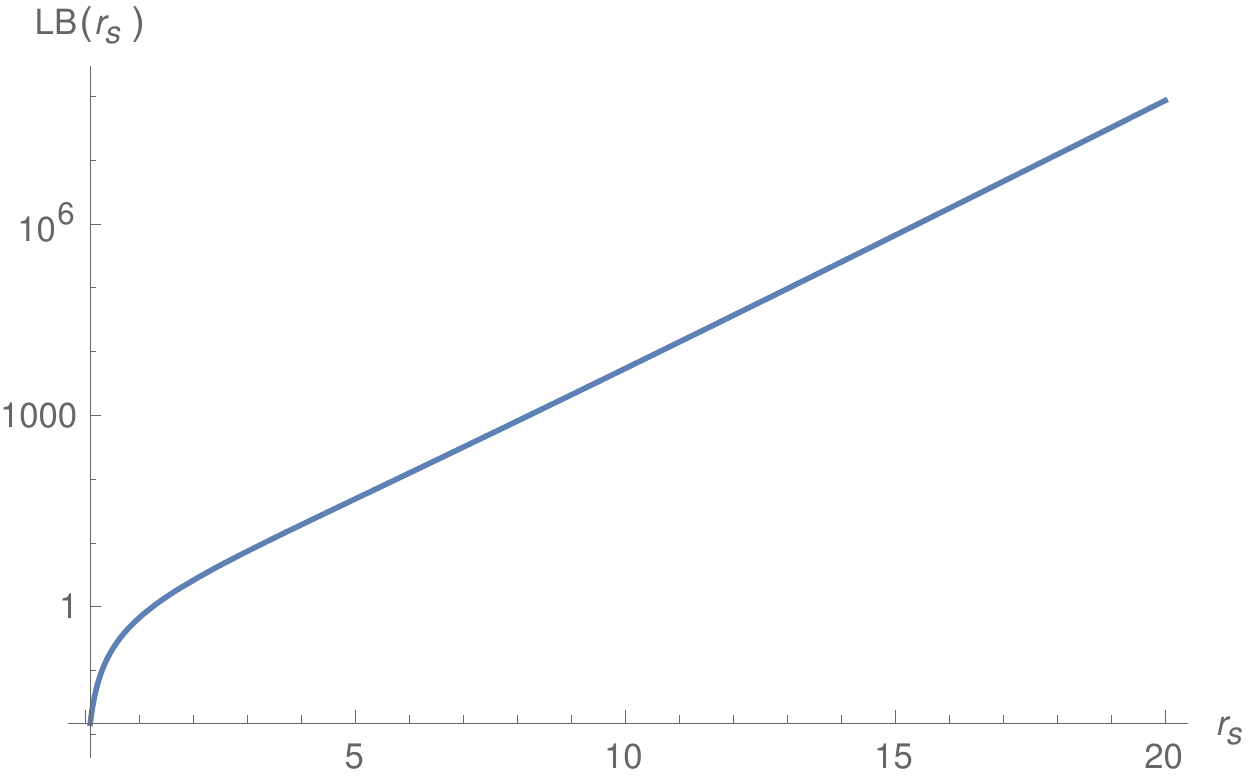}
  \caption{ $LB(r)$ plotted against $r$ for $l = 1$. We see that the lower bound goes exponentially as $r$ at larger values of $r$.   }
  \label{search_time}
  \end{figure}

This is plotted in fig.\ref{search_time}. One can see that for large times the $LB(r)$ increases exponentially with $r$.  This would be the lower bound even if $\alpha \neq 0$, because $\alpha$ only adds randomness to the search and hence would increase the  search times. 

  \emph{Generic lower bound:-}
 We can simply use the fact that the probability distribution evaluated by Baye's theorem is not as concentrated near the source as $S(x,y)$ to simply evaluate a lower bound on search time as follows. First let us assume that the searcher knows that the source is located at the origin with a probability $1$. Then the smallest time taken by the searcher to reach the source goes as $r$, the distance between the source and the searcher. In case the searcher instead knowns that the source is located at two points $\vec{x_1}$ and $\vec{x_2}$  with probability $p_1$ and $p_2$. Then, the smallest possible search time would simply go as $ p_1 | \vec{x_1} - \vec{x_s}| +  p_2 | \vec{x_2} - \vec{x_s}| $ where $\vec{x_s}$ is the searchers position. This is obvious because out of $N$ possible measurements, the source is seen at $Np_1$ times at $\vec{x_1}$ and $Np_2$ times at $\vec{x_2}$. One could extend this to say that for a source probability distribution $P(\vec{x})$ as understood by the searcher, the shortest time to reach the source on an average should go as $\int d\vec{x} |\vec{x_s}-\vec{x}| P(\vec{x}) $ .

Since the fact that the probability distribution evaluated by Baye's theorem is not as concentrated near the source as $S(\vec{x})$, the search time evaluated using any strategy that utilizes the probability distribution as measured by a searcher could never be smaller than $ \frac{1}{v_s} \int d\vec{x} |\vec{x_s}-\vec{x}| S(\vec{x}) $ ($v_s$ is the speed of the searcher, which we take to be equal to $1$ below), which for $S \sim K_0(r/l)$ is

\begin{eqnarray}
LB(r_s)&\sim & \int r d\theta dr  K_0(r/l) \sqrt{ (r_s - r\cos\theta)^2 + r^2 \sin^2\theta} \nonumber \\
 &= & \int r d\theta dr  K_0(r/l) \sqrt{ r_s^2  + r^2  -2 r r_s \cos \theta} \nonumber \\
 \label{lower bound} 
\end{eqnarray}

Now since
\begin{eqnarray}
\frac{1}{ \sqrt{ r_s^2  + r^2  -2 r r_s \cos \theta} }&=&    \sum_{l=0,\infty} \frac{r^l}{r_s^{l+1}}P_l(\cos \theta) ,\; r_s > r\nonumber \\
 &=& \sum_{l=0,\infty} \frac{r_s^l}{r^{l+1}}P_l(\cos \theta) ,\; r_s < r \nonumber \\
\end{eqnarray}
   implies
   \begin{eqnarray}
\frac{1}{2r_s - 2 r \cos \theta }\frac{d}{dr_s }  \sqrt{ r_s^2  + r^2  -2 r r_s \cos \theta}  &=&     \sum_{l=0,\infty} \frac{r^l}{r_s^{l+1}}P_l(\cos \theta) ,\; r_s > r\nonumber \\
 &=&  \sum_{l=0,\infty} \frac{r_s^l}{r^{l+1}}P_l(\cos \theta) ,\; r_s < r \nonumber \\
\end{eqnarray}
   
\begin{figure}
  \includegraphics[scale=.75]{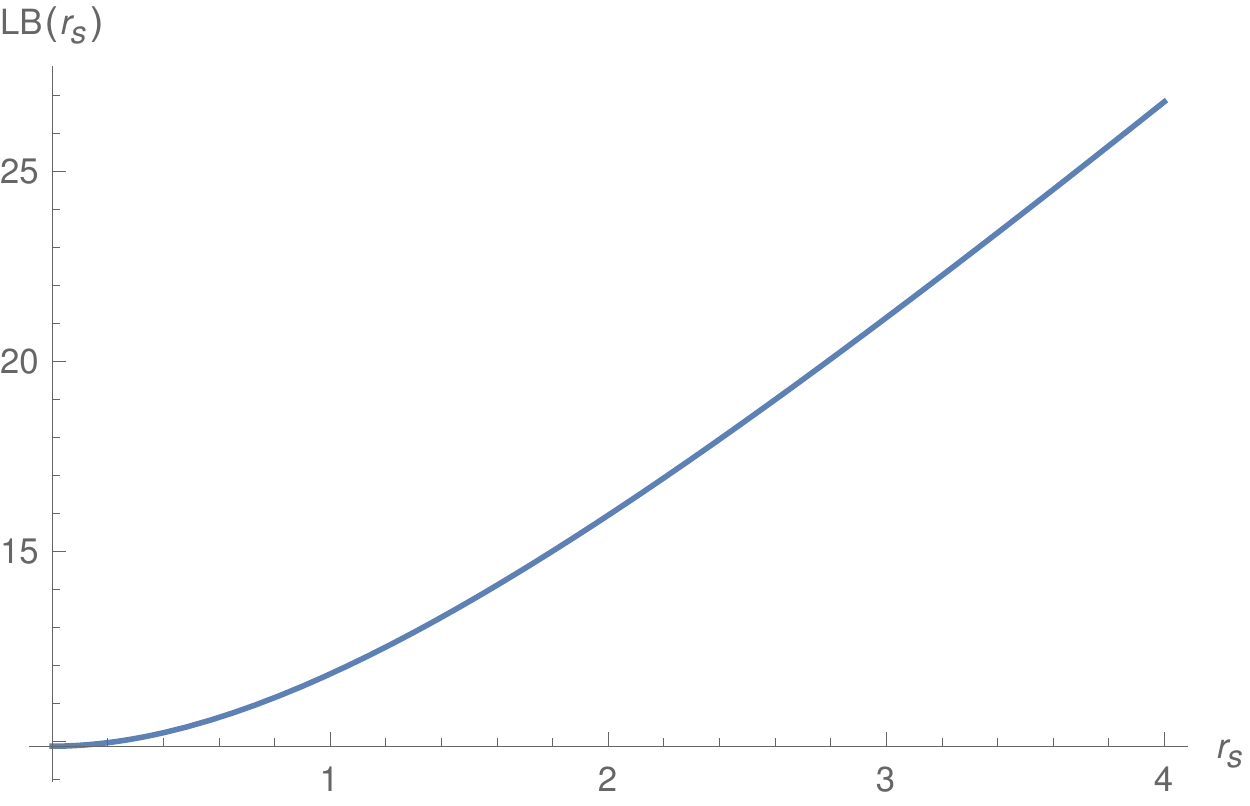}
  \caption{  $LB(r_s)$ plotted as a function of $r_s$ solving Eq.\ref{lower bound} with $l=1$. As can be seen at large values of $r_s > l$ we have an expected lower bound going as $r_s$. Also note that for $r_s = 0$ the lower bound on the search time    is not zero.  }
  \label{lower_bound}
  \end{figure}

    Hence
    \begin{eqnarray}
 \frac{d LB(r_s)}{dr_s}&= & \int_0^{r_s} \int_0^{2\pi} r d\theta dr  K_0(r/l) \sum_{l=0,\infty} \frac{2( r_s -  r \cos \theta ) r^l}{r_s^{l +1 }}\nonumber\\
 &&P_l(\cos \theta) + \int_{r_s}^\infty \int_0^{2\pi} r d\theta dr \nonumber \\
 && K_0(r/l) \sum_{l=0,\infty} \frac{2 ( r_s -  r \cos \theta )r_s^{l}}{r^{l+1}}P_l(\cos \theta)\nonumber \\
 \label{total}
\end{eqnarray}
  For large values of $r_s$, the second integral would contribute minisculely. Also majority contribution in first term would only show from the $l=0$. Hence
 
   \begin{eqnarray}
  \frac{d LB(r_s)}{dr_s} &\approx & 2 \times 2 \pi  \int_0^{r_s} r   dr  K_0(r/l) \times \frac{ r_s}{r_s }   \nonumber \\ 
 &\approx &  4\pi l^2     \nonumber \\ 
 \label{slope}
\end{eqnarray}
As $r_s$ is made smaller, other contributions start appearing. However, we note that as $r_s$ becomes larger and larger, the lower bound on search time goes simply as $r_s$. This simply states the fact that as $r_s$ becomes large, the range over which the region of size $l$ surrounding the source looks like a point object to the searcher. This behavior is seen by solving Eq.\ref{lower bound}  for $l=1$ as plotted in fig. \ref{lower_bound}. 

For small values of $r_s$ one  could simply expand
 \begin{eqnarray}
  LB(r_s)&\sim & \int r d\theta dr  K_0(r/l) \sqrt{  r_s^2 + r^2 -2r r_s \cos \theta} \nonumber \\
 &= & \int r^2  d\theta dr  K_0(r/l) \sqrt{  1 +\frac{r_s^2}{r^2} -2  \frac{r_s}{r} \cos \theta} \nonumber \\
  &= & \int_0^{\infty} \int_0^{2\pi}   d\theta dr  K_0(r/l) ( \frac{r_s^2}{2}+ r^2 -2^2  \frac{r_s^2}{8 } \cos^2 \theta) \nonumber \\
   &= & \int_0^{\infty} \int_0^{2\pi}   d\theta dr  K_0(r/l) ( \frac{r_s^2 \sin^2 \theta }{2}+ r^2 ) \nonumber \\
  &=& 2\pi  \frac{\pi}{2} l^3   + \frac{1}{2}\pi  \frac{\pi l}{2}    r_s^2\nonumber \\
&=& \pi^2 l^3 + .25 \pi^2 l r_s^2
  \end{eqnarray}
which is the behavior for $r_s << l$. 
 Note that 
 
 \begin{eqnarray}
LB(r_s)&\sim & \int r d\theta dr  K_0(r/l) \sqrt{ (r_s - r\cos\theta)^2 + r^2 \sin^2\theta} \nonumber \\
 &= &l^3 \int \frac{r}{l} d\theta d\frac{r}{l}  K_0(r/l) \sqrt{ (\frac{r_s}{l})^2  + (\frac{r}{l})^2  -2 \frac{r}{l} \frac{r_s}{l} \cos \theta} \nonumber \\
 &= &l^3 \int x d\theta dx  K_0(x) \sqrt{ (\frac{r_s}{l})^2  + x^2  -2 x \frac{r_s}{l} \cos \theta} \nonumber \\
\end{eqnarray} 
 Hence all that matters is how $r_s$ compares to $l$. From fig.\ref{lower_bound} we can see that when $r_s > 2l$, the behavior of $LB(r_s)$ is linear. From Eq.\ref{slope} we can see that slope of this line is  $4 \pi l^2$. 
  One can hence say that 
 \begin{eqnarray}
 LB(r_s)> \frac{\pi^2 l^3}{2} + 4 \pi  l^2 r_s
 \end{eqnarray}

 If we instead consider the lower bound on search time given a particular entropy of the source  probability distribution we have to minimize
 
 \begin{eqnarray}
 LB &=& \int 2 \pi r dr d\theta r S(r,\theta) - \lambda \int 2\pi r dr d\theta (S(r,\theta) \ln S(r,\theta) +E)\nonumber\\
 && -\beta (\int 2\pi r dr d\theta S(r,\theta) - 1) \nonumber \\
 \end{eqnarray}
We have assumed the searcher is located at $r=0$. Here $\lambda$ is the Lagrange multiplier that sets $E$ to the entropy of the probability distribution $S(r,\theta)$.  Minimizing   w.r.t $S(r,\theta)$ gives
 \begin{eqnarray}
 r   -\lambda (\ln S(r,\theta) +1) - \beta &=& 0 \nonumber \\
 \end{eqnarray}
 which solves to
 \begin{eqnarray}
 S(r,\theta) = e^{r/ \lambda-\beta / \lambda - 1}
 \end{eqnarray}
 $\lambda <0$ for consistency. 
 Requiring that 
 \begin{eqnarray}
  \int 2\pi r dr d\theta (S(r,\theta) \ln S(r,\theta)  ) &=& - E \nonumber \\ \rightarrow  (2\pi)^2 \lambda     e^{\frac{\beta }{\lambda }-1} (\beta -3 \lambda ) = - E\nonumber\\
 \int 2\pi r dr d\theta S(r,\theta)   &=& 1 \rightarrow  e^{  \beta / \lambda -1} (2\pi)^2 \lambda^2  = 1 \nonumber \\
 \end{eqnarray}
 which implies $ (\beta - 3\lambda) = -E \lambda \rightarrow (\frac{\beta}{\lambda } ) = 3- E$ and $ \lambda =- \frac{1}{2\pi } e^{E/2 -1}$
 
 Hence the lower bound is 
 \begin{eqnarray}
 LB &=& \int 2 \pi r dr d\theta r S(r,\theta) = -2(2\pi)^2  \lambda ^3 e^{\frac{\beta }{\lambda }-1} = -2\lambda = \frac{e^{E/2-1 }}{\pi}\nonumber \\
 \end{eqnarray}

\emph{Conclusion:- }
In \cite{infotaxis} the difficulty in evaluating the search time for Infotaxis was highlighted.  and instead   a calculation for a different search strategy, which does not utilize information about past hits, was presented. They evaluated the lower limit for  search time for this strategy in  certain limits as  $\sim e^{ E}$, where $-E$ is the entropy of the probability distribution of finding the source. We however have in this work we evaluated a lower bound on   the average search time in  a search strategy that seeks to evaluate the probability distribution of finding the source, given the information of past hits, such that rate of jumps to a neibhouring site is proportional to the normalized difference of evaluated probability of finding the source with the present site of the searcher. This lowerbound goes as the exponential of distance from the source for large distances. We then provided an expression for the lower bound for the search time for any cues based search strategy.     For a general probability distribution with entropy -$E$, we showed that the lowerbound goes as $e^{E/2}$. We see that the lowerbound again goes as $e^{E/2}$ which is similar to $e^{E}$ in ref.\cite{infotaxis}, which was evaluated for a non cue based search strategy for  the limit in which the search time as well as entropy are much larger than $1$. The similarity almost begs  a conjecture that for probability distributions that do not narrow the source position well, cue based searches do not perform appreciably better than non cue based searches. It would be interesting to further explore this statement through further research.

	 \emph{Appendix:-}
	 To simplify things, let us consider the system in one dimension. The final result can be easily generalized to higher dimensions.  
	  We have

	\begin{eqnarray}
	T(x)&=&- dt + T(x+dx)[\beta   \Pi(x+dx) +\alpha \Delta(x+dx)  ]   \nonumber \\
	&+& T(x-dx)  [\beta\Pi(x-dx) +\alpha \Delta(x-dx) ] \nonumber \\
 &+& T(x) [1-    \beta   (  \Pi(x+dx)^- + \Pi(x-dx)^-   +2\alpha \Delta(x)) ]  \nonumber \\
 \label{eq}
		\end{eqnarray}
		where 
		\begin{eqnarray}
		\Pi(i) &=& \Theta(P(x)-P(i) ) \frac{ (P(x)-P(i) )}{P(x)} \nonumber \\
		\Pi(i)^- &=& \Theta(-P(x) + P(i) ) \frac{ (-P(x)+P(i) )}{P(x)} \nonumber \\
		\Delta(i) &=& 1 \quad P(x) = P(i)\nonumber\\
		&=& 0 \quad P(x) \neq P(i) \nonumber \\
		\end{eqnarray}
		The eq. \ref{eq} simply states that we can reach point $x$ from any of its neighbors $x+dx$ and $x-dx$, which subtracts time $dt$ from times $T(x+dx)$ , $T(x-dx)$ to reach source from these sites. 		Each of the times $T(x+dx)$ , $T(x-dx)$ are multiplied by the probabilities to make the jump from $x+dx$ and $x-dx$ to $x$ respectively. The term multiplying $T(x)$ on the RHS is simply the probability of not making a jump to the neighbors $x+dx$,$x-dx$. $\alpha$ is the probability of making a jump randomly in case the neibhouring site has the same probability of finding the source as present site. 
		
	This eq. \ref{eq} becomes
	\begin{eqnarray}
	0&=&- dt +  \beta  \;[T(x+dx)   \Pi(x+dx)   +  
	  T(x-dx)   \Pi(x-dx) ] \nonumber \\
 &-& T(x) \beta    [     (  \Pi(x+dx)^- + \Pi(x-dx)^-  ) ] +\beta \alpha   dx^2 \nabla^2 T(x) \nonumber \\
		\end{eqnarray}
		or
		\begin{eqnarray}
	0&=& -dt +   \beta  \;[ (T(x ) + dx \partial_x T(x) )   \Pi(x+dx)   \nonumber \\& +&
	      (T(x ) - dx \partial_x T(x) )  \Pi(x-dx) ] \nonumber \\
 &-& T(x) \beta    [     (  \Pi(x+dx)^- + \Pi(x-dx)^-  ) ] +\beta \alpha   dx^2 \nabla^2 T(x)  \nonumber \\
		\end{eqnarray}

		or

	\begin{eqnarray}
	0&=& -dt -    \beta   \; dx^2 \; \partial_x T(x)[  \frac{\partial_x P(x)}{P(x)}   \Theta(P(x)-P(x+dx)) \nonumber \\
	&+&  \frac{ \partial_x P(x)}{P(x)}   \Theta(P(x)-P(x-dx)) ] \nonumber \\
  &+& T(x) \beta    [     (  \Pi(x+dx) + \Pi(x-dx)  ) ]  \nonumber \\
 &-& T(x) \beta   [     (  \Pi(x+dx)^- + \Pi(x-dx)^-  ) ]  +\beta \alpha   dx^2 \nabla^2 T(x) \nonumber \\
		\end{eqnarray}

	now
		\begin{eqnarray}
	\Pi(i)- \Pi(i)^- &=&[ \Theta(P(x)-P(i) ) +  \Theta(-P(x) + P(i) ) )]\nonumber \\
	&& \frac{ (P(x)-P(i) ) }{P(x)} 
	=  \frac{(P(x)-P(i) ) }{P(x)}  \nonumber \\
	\end{eqnarray}
	hence
		\begin{eqnarray}
	0&=&-  dt - \beta   \; dx^2 \; \partial_x T(x) \frac{ \partial_x  P(x)}{P(x)} \nonumber \\
	&-& T(x) \beta    [   \frac{  (  P(x+dx) + P(x-dx)-2P(x  ) }{P(x)}]  +\beta \alpha   dx^2  \nabla^2 T(x) \nonumber \\
	\end{eqnarray}	
	or
		\begin{eqnarray}
	0&=& -  P(x) dt - \beta   \; dx^2 \; \partial_x T(x) \partial_x  P(x) \nonumber \\
	&-& T(x) \beta     \; dx^2 \;  \nabla^2 P(x)  +\beta \alpha  dx^2 P(x) \nabla^2 T(x) \nonumber \\
	\end{eqnarray}	
	 which becomes in higher dimensions
 	 	
	 	\begin{eqnarray}
	 0&=& -P(x)  - \beta \nabla T(x) \cdot \nabla P(x) \nonumber \\
	 &-& \beta T(x)    \nabla^2 P(x) +\beta \alpha  P(x)  \nabla^2 T(x)  \nonumber \\
	 \end{eqnarray}	
	 we have redefined  $\frac{\beta     dx^2}{dt} \rightarrow \beta$ above.

\end{document}